# Evidence that Cross-Domain Re-interpretations of Creative Ideas are Recognizable

**Apara Ranjan, Liane Gabora, and Brian O'Connor**

Department of Psychology
University of British Columbia
Okanagan campus, 3333 University Way
Kelowna BC, V1V 1V7, CANADA
apara.ranjan@ubc.ca, liane.gabora@ubc.ca, brian.oconnor@ubc.ca

**Abstract**

The goal of this study was to investigate the translate-ability of creative works into other domains. We tested whether people were able to recognize which works of art were inspired by which pieces of music. Three expert painters created four paintings, each of which was the artist's interpretation of one of four different pieces of instrumental music. Participants were able to identify which paintings were inspired by which pieces of music at statistically significant above-chance levels. The findings support the hypothesis that creative ideas can exist in an at least somewhat domain-independent state of potentiality and become more well-defined as they are actualized in accordance with the constraints of a particular domain.

## Introduction

Where does the uniqueness of a creative work come from? The creative process involves inspiration, translation, and re-interpretation, as well as the accessing and combining of knowledge, experiences, and ideas, to come up with new creative outputs (Cropley, 1999; Feldhusen, 1995, 2002; Munford & Gustafson, 1988; Sternberg & Lubart, 1995). One source of the uniqueness of a creative work, at least with respect to creativity in the arts, appears to be the personal style of the creator. It was shown that an individual's creative outputs exhibit an identifiable character that is recognizable not just within but also across domains (Gabora, O'Connor, & Ranjan, 2012). For example, creative writing students familiar with each other's writing (and not with each others art) guessed at above chance levels which fellow student had created which piece of art. Thus the uniqueness of a creative work derives, at least in part, from the personal style of the creator.

However, it is likely that the uniqueness of a particular creative work has other sources as well. Creative inspiration may come from a work in same domain as the work it inspires (as when one poem inspires an idea for another poem, and sometimes it is in a different domain as when a piece of music inspires a poem). New creative products often have a striking similarity to the known domain instances (Ward, 1994). Thus, in addition to the creative style of the creator himself or herself, another source of the uniqueness of a particular creative work might be one or more elements of the world that inspired the creator to initiate the creative work.

There are several phenomena suggesting that the creative work need not even be in the same domain as the inspirational source for the work, as in for example, an artist when an artist is so moved by a poem or film that that poem or film inspires him or her to create a painting. In the late nineteenth century, musical terminology was used as titles for the paintings. Another practice of this period was to associate particular kinds of music with particular colors. Music also frequently served as a direct inspiration for paintings, as it continues to do today. We now review phenomena that point to cross-domain interpretation of ideas as a source of the character of creative works: synesthesia, the practice of ekphrastic expression, and cross-domain style.

### Synesthesia

Individuals referred to as synesthetes naturally and spontaneously translate stimuli into another sensory domain. For example, they may see particular letters and numbers in particular colors. Ramachandran (2003) proposes that synesthesia occurs as a result of hyper-connectivity in the brain and partial collapse of the barrier between sensory domains. Artists, poets, and novelists, are more likely than average to be synesthetes, which suggests that synesthetically driven re-interpretation of inputs from one modality to another may play some role in these creative domains (Ramachandran, 2003).

## Ekphrastic Expression

There is a tradition in the arts of interpreting art from one medium (*e.g.*, oil paint) into another (*e.g.*, watercolor) and thereby coming to know its underlying essence. This practice is referred to as *ekphrastic expression*. The idea behind ekphrastic expression is that an artist may have a more direct impact on an audience by translating one medium of art into another medium because this involves capturing, and thereby becoming intimate with, its underlying form or essence. Modern day film composers attempt to compose music that conveys the emotional tone of the events portrayed in a film, thereby heightening the viewer's experience of these events. Thus film scoring can be seen as a form of ekphrasis. The application of ekphrastic methods in the arts supports the idea that creative individuals extract patterns of information from the constraints of the domains in which they were originally expressed, and transform them into other domains.

## Cross-Media Style

Another reason to suspect that the character of creative works arises through cross-domain re-interpretations of creative ideas is the widespread phenomenon of *cross-media style*. This refers to artistic style that is demonstrated by works of art in more than one medium. For example, the term rococo is applied to the painting, sculpture, literature, and music of a certain period. These forms are considered to be abstract archetypal forms or potentialities that make the mind interested in certain processes or arrangements in the artwork (Burke, 1957). Evidence for cross-media style shows how works of art in different media could be similar in terms of psychophysical, collative, and ecological properties (Hasenfus, 1978). There is also a pattern in the choice of elements (*i.e.*, colors, shapes, words, or instruments) made by an artist to create a novel piece of music, painting or writing (Berlyne, 1971).

## Aims of the Present Research

The goal of the current research was to test the hypothesis that the uniqueness of a creative work derives partly from conceptual entities themselves which get abstracted and reinterpreted into different concrete forms by investigating whether it is possible to recognize the source of inspiration for a creative work when that source of inspiration comes from a different domain. In other words, we aimed to determine whether it is possible to identify a work of art that exists in one medium after it has been recreated in another medium.

# Methods

This study examined whether people were able to correctly recognize which works of art were inspired by which pieces of music. The study was divided into two phases. In the first phase, artists created paintings in response to musical stimuli. In the second phase, naïve participants attempted to determine which piece of music was used as the source of inspiration for each artwork.

### Phase One

In the first phase of the study, expert artists were asked to create four paintings, each of which was the artist's interpretation of one of four different pieces of instrumental music.

*Participants*. Three local expert artists, each with approximately 25 years of experience in the field of painting, were recruited for this study. They received 50$ each for their participation.

*Musical Stimuli*. Four pieces of piano music were used in this study. The pieces were selected from a pool of 45 pieces chosen as exemplars of different musical styles: baroque classical, romantic, jazz and contemporary from commercially produced sound track CDs. Each of these original 45 pieces of music was cropped to three minutes duration, and then rated by three raters on 64 descriptive adjectives on five point Likert scales. The adjectives were derived from previous research on the collative properties of stimulus patterns, from measures of affective reactions to artwork (Berlyne, 1974), and from the affective circumplex (Russel, 1980; Watson & Tellegen, 1985). The raters had no previous training in music.

Factor analysis and multidimensional scaling were used to compute the basic dimensions of aesthetic experience in the ratings, and reveal how the 45 pieces of music were dispersed in the dimensional spaces. The Euclidean distances between the pieces of music across the spaces were used to select four pieces of music from different regions that were clearly dissimilar from each other. The four pieces of music selected were the following:

(1) 'Love is a Mystery' composed by Ludovico Einaudi

(2) Number 29 B Flat Major', by Ludwig van Beethoven

(3) 'Circus Gallop' by Marc-André Hamelin

(4) 'All of Me' by Jon Schmidt

*Creation of Artworks*. Each of the three artists created one painting for each of the four pieces of piano music, for a total of 12 paintings. On the day that a painting was to be created, the artists were provided with a single piece of music. They were asked to reinterpret each piece of music as a painting, *i.e.*, to paint imagining what the music would look like if it were a painting. They were instructed to paint while listening to the music, and encouraged to listen to the music as many times as they wished while they painted. They were asked to paint using the materials they thought could most effectively express the music (*e.g.*, watercolors, oils, acrylic, or chalk). All participants were instructed to take up to a maximum of 120 minutes to listen to the music and complete the painting.

The paintings were created in the artists' personal studios. The artists were instructed to complete their painting at one sitting without interruption until they were finished. In order to limit the influence of the previous piece of music on the new painting, artists were instructed not to re-listen to the piece of music after the painting was finished, and there was a gap of four days between each painting session.

Representative examples of the music-inspired paintings obtained in Phase One of the study are provided in Figures 1 and 2. These paintings constituted the stimuli that were used in Phase Two.

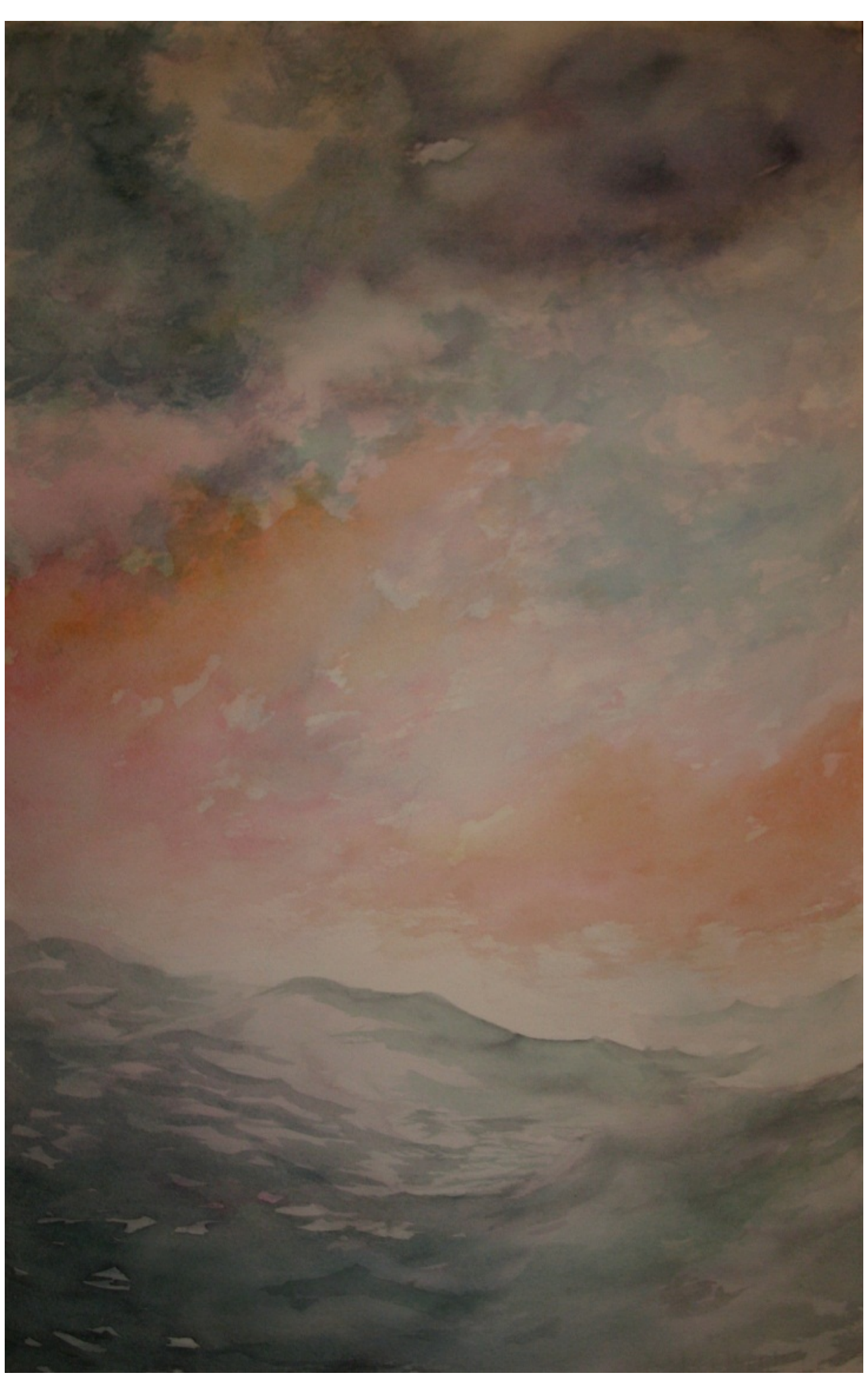

*Figure 1. A painting created by one of the artists in response to the piece Number 29 B flat major' by Ludwig van Beethoven.*

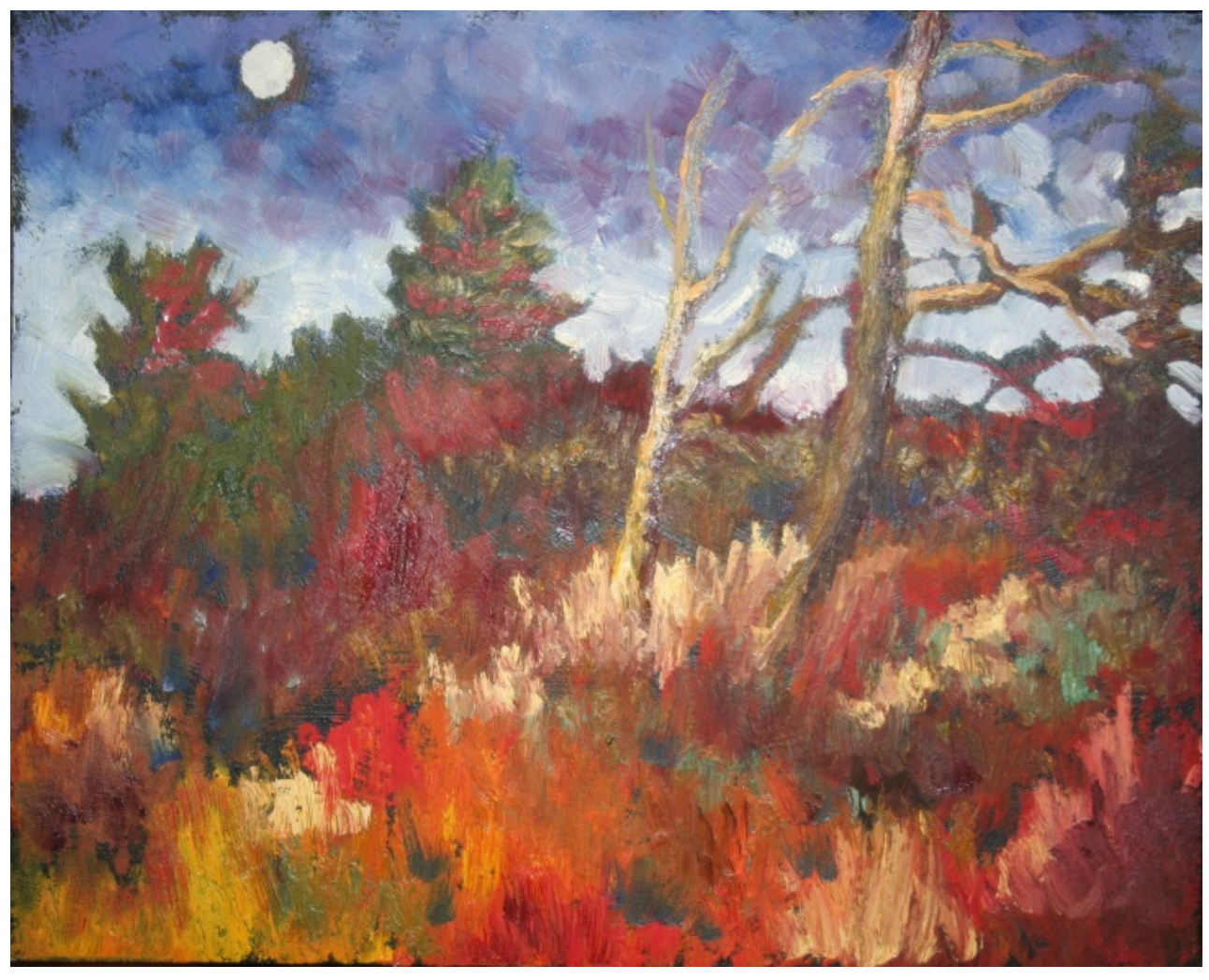

*Figure 2. A painting generated by one of the artists as an interpretation of the piece 'All of Me' composed by Jon Schmidt.*

**Phase Two.**

In the second phase of the study we tested whether people are able to recognize which pieces of music were interpreted as which paintings.

*Participants.* 52 undergraduates enrolled in psychology courses at the University of British Columbia participated in the study using the SONA system which matches research projects to potential particiaptns. They received partial course credit for their participation.

*Procedure and Materials.* This part of the study was set up online. The four pieces of music and the twelve paintings from Phase One were used. Each painting was displayed on a different webpage, which participants looked at one at a time. In addition to the painting, each webpage had links to the four pieces of music. Participants were asked to look at the painting, to listen to the four pieces of music, and identify which piece of music they think was most likely to have inspired the painting that appeared on that page.

*Analytic Methods.* Two statistics, *H* and *Hu*, were computed to assess the accuracies of the participants' paintings-to-music matches. *H* is the simple hit rate, or the proportion of correct guesses. The *Hu* was based on signal detection theory research and it corrects for chance guessing and for response bias, such as the tendency to use particular response categories more or less than other response categories (Wagner, 1993). *H* and *Hu* were computed for each participant. One-sample t-tests and a data randomization procedure (Manly, 2007) were then used to assess the statistical significance of the mean *H* and *Hu* values. This was accomplished by determining whether the mean *H* and *Hu* values were significantly different from the *H* and *Hu* values that would have been obtained had participants provided random guesses.

## Results

The results of this study are summarized in Table 1. The mean hit rates were $H$ = .51 and $Hu$ = .54. The mean hit rates that would have been obtained on the basis of random guesses for these questions were .25 and .25, respectively. Both hit rates were statistically significant according to both conventional and data randomization t-tests ($t$ (51) = 6.9, $p$ < .001, and $t$ (51) = 7.2, p < .001). The $r$ effect sizes were large, .70 and .71. Thus participants were able to identify at above-chance levels which paintings were inspired by which pieces of music.

| | **Mean Hit Rate** | **Chance Hit Rate** | ***t (df)*** | ***r* Effect Size** |
|---|---|---|---|---|
| Hit Rate (*H*) | .51 | .25 | 6.9 (51) | .70 |
| Unbiased Hit Rate (*Hu*) | .54 | .25 | 7.2 (51) | .71 |

Table 1: Table for Mean Hit Rates, *t*-Test values, and *r* Effect Size for identification of paintings inspired by the four pieces of music. All hit rates and t values were statistically significant at the .05 level.

## Discussion

To our knowledge, no previous study has tested the hypothesis that the core idea behind a creative work is recognizable when it is translated from one domain to another. This hypothesis was supported by our finding that when pieces of music were re-interpreted as paintings, naïve participants were able to guess significantly above chance which piece of music inspired which painting. Although the medium of expression is different, something of its essence of the idea remains sufficiently intact for an observer to detect a resemblance between the new work and the source that inspired it.

There is a longstanding a debate concerning the extent to which the semantic complexity of artistic works is amenable to scientific methods (Becker, 1982). We suggest that at their core, creative ideas may be much less domain-dependent than they are generally assumed to be, and that indeed it is possible for their domain-specific aspects to be stripped away such that they exist in an abstracted state of potentiality and being re-expressed in another form. This interpretation of the results is consistent with a number of other phenomena, such as synesthesia and ekphrastic expression, discussed in the Introduction. The research reported on here may be a step toward distinguishing between domain-specific and domain-general aspects of creative works. The notion that the uniqueness of a creative work derives at least in part from, not just the personal style of the creator, but from encounters with works in domains that differ from the domain of the creative output, or even different kinds of experiences altogether, suggests that creative works emerge from a state of potentiality that can manifest in different domains. A creative idea can exist in form that is freed of the constraints of a particular domain, and that the creator's job may be in part to, to simply allow that domain-independent entity to take a particular form, using domain-specific expertise and the tools of his or her trade. Over time they may become more fully actualized, and well-defined, as they are considered from different perspectives in accordance with the constraints of the domain in which they are expressed. It was suggested that this is because an individual's creative outputs are expressions of a particular underlying uniquely structured self-organizing internal model of the world, or worldview. The proposal that creativity reflects the tendency of a worldview to transform in such a way as to find connections, reduce dissonance, and achieve a more stable structure. When a work is translated from one domain (*e.g.*, music) into another (*e.g.*, painting), the two works will be recognizably related because they are expressions of the same core idea. This view of creativity is consistent with previous research showing that midway through a creative process, an idea may exist in a state of potentiality in a 'half-baked' state of potentiality, in which one or more elements are ill-defined (Gabora, 2001, 2005, 2010, 2011; Gabora & Holmes, 2010; Gabora & Saab, 2011; Schwartz & Gabora, submitted).

Although that idea that at least some creative tasks involve the abstraction of form from one domain and re-expression of it into another domain seems intuitively obvious to many of the artists we have spoken with, it stands in contrast with most academic theories of creativity. Creativity is typically portrayed as a process of searching and selecting amongst candidate ideas that exist in discrete, well-defined states. This can be traced back to early views arising in the artificial intelligence community, wherein creativity was thought to proceed by heuristically guided search through a space of possible solutions (Newell, Shaw & Simon, 1957; Newell & Simon, 1972; Simon, 1973, 1986) or possible problem representations (*e.g.,* Kaplan & Simon, 1990, Ohlsson, 1992). The view that creativity proceeds through a process of search and selection also assumed in more contemporary theories, such as the theory that creativity is a Darwinian process; *i.e.*, new ideas are obtained by generating variations more or less at random and selecting the best (*e.g.,* Campbell, 1965; Simonton 1999a,b, 2005).

The results reported here inevitably lead to the question of what it was about the paintings that made it possible to trace them to the artworks that had inspired them. Aesthetic perception stimulated by works of art, such as paintings, drawings, or sculptures, may lead to complex responses. These responses could be emotional, cognitive, behavioral, and/or physiological in nature, and amenable to re-expression in another form. One theory that may provide insight into the mechanisms underlying cross-domain translation of ideas is *common coding*. According to this theory, information in different brain regions is represented using the same basic coding system, thereby facilitating the sharing and re-interpretation of their contents. This is said to be particularly the case for three kinds of representations: motor representations, perceptual representations, and the covert activation of motor and perceptual representations that occurs when we imagine movements. Any one of these movement representations can automatically trigger the other two (Prinz, 1997). The common coding hypothesis could thus also at least partially account for the ability to re-interpret ideas across different domains, and to recognize the essence of the idea after it has been re-interpreted. Moreover, it allows for the possibility of recognizing dynamic inspirational sources (such as music) from static traces (such as paintings).

Research by Feedberg and Gallese (2007) on the perceiving action in the artwork provides another possible clue to the mechanisms underlying cross-domain interpretation of creative ideas. They propose that art observers implicitly imitate the creative actions undertaken by the artist in the making of the work. In our study it is possible that observers were not just perceiving action in art but were also able to match it rhythm and tempo of the music. This phenomenon of action perception in paintings could also at least partially account for the ability to recognize the essence of ideas interpreted across domains. In order to recognize the inspiration of an artwork or a cross-media style, expertise in a domain might stimulate the action system while the observer imagines how the artwork was created.

Future research will focus on the role of expertise in the recognition of a connection between works in different domains. We hypothesis that expertise in a domain might increase the activation of the action system while the observer imagines how the artwork got created, thereby enhancing the capacity for recognition of cross-domain re-interpretation in a task such as this.

## Acknowledgments

We are grateful for funding to the second author from the Natural Sciences and Engineering Research Council of Canada and the Concerted Research Program of the Flemish Government of Belgium. We thank Jon Corbett for his assistance in carrying out this research. We also thank the editor and anonymous reviewers for comments and suggestions.